\documentclass[fleqn,11pt]{article}
\usepackage{amsfonts,epsfig}
\usepackage{times}
\usepackage[vcentermath,enableskew]{youngtab}

\newcommand{\Ga}{\alpha}

\newcommand{\Gd}{\delta}
\newcommand{\Ge}{\epsilon}

\newcommand{\Gg}{\gamma}
\newcommand{\GG}{\Gamma}

\newcommand{\Go}{\omega}

\newcommand{\Gth}{\theta}
\newcommand{\GTh}{\Theta}

\newcommand{\cA}{{\scriptscriptstyle\cal A}}
\newcommand{\cB}{{\scriptscriptstyle\cal B}}
\newcommand{\cC}{{\scriptscriptstyle\cal C}}
\newcommand{\cD}{{\scriptscriptstyle\cal D}}

\newcommand{\cM}{{\scriptscriptstyle\cal M}}
\newcommand{\cN}{{\scriptscriptstyle\cal N}}
\newcommand{\cK}{{\scriptscriptstyle\cal K}}
\newcommand{\cL}{{\scriptscriptstyle\cal L}}
\newcommand{\cP}{{\scriptscriptstyle\cal P}}

\newcommand{\CA}{{\cal A}}
\newcommand{\CB}{{\cal B}}

\newcommand{\CM}{{\cal M}}
\newcommand{\CN}{{\cal N}}

\newcommand{\CL}{{\cal L}}
\newcommand{\CP}{{\cal P}}
\newcommand{\CQ}{{\cal Q}}

\newcommand{\cV}{{\cal V}}

\newcommand{\dA}{{\dot{A}}}
\newcommand{\dB}{{\dot{B}}}

\newcommand{\scT}{{\scriptscriptstyle T}}

\newcommand{\indA}{{\scriptstyle A}}
\newcommand{\indB}{{\scriptstyle B}}
\newcommand{\inddA}{{\scriptstyle \dA}}
\newcommand{\inddB}{{\scriptstyle \dB}}

\newcommand{\Beps}{\overline{\Ge}}
\newcommand{\Bchi}{\overline{\chi}}
\newcommand{\Bpsi}{\overline{\psi}}

\newcommand{\ft}[2]{{\textstyle {\frac{#1}{#2}} }}

\newcommand{\dd}{\partial}

\newcommand{\ra}{\rightarrow}
\newcommand{\I}{{ i}}

\newcommand{\be}{\begin{equation}}
\newcommand{\ee}{\end{equation}}
\newcommand{\ben}{\begin{displaymath}}
\newcommand{\een}{\end{displaymath}}
\newcommand{\ba}{\begin{eqnarray}}
\newcommand{\ea}{\end{eqnarray}}
\newcommand{\nn}{\nonumber}
\newcommand{\non}{\nonumber\\}

\newcommand{\mathon}{\mathversion{bold}}
\newcommand{\mathoff}{\mathversion{normal}}

\newcommand{\la}{\label}

\newcommand{\Ref}[1]{(\ref{#1})}

\newcommand{\vl}{{\vphantom{[}}}
\newcommand{\VV}[2]{{\cV^{\,#1}{}\!^\vl_{#2}}}
\newcommand{\TT}[2]{{T^\vl_{#1|#2}}}

\newcommand{\mfg}{{\mathfrak g}}

\newcommand{\ovi}{{\overline{i}}}
\newcommand{\ovj}{{\overline{j}}}
\newcommand{\ovk}{{\overline{k}}}
\newcommand{\ovl}{{\overline{l}}}

\newcommand{\cro}{\!\times\!}
\newcommand{\equ}{\!=\!}
\newcommand{\pls}{\!+\!}
\newcommand{\mis}{\!-\!}

\newcommand{\vecso}{$({\bf 8}${\mathon$+$\mathoff}${\bf n})$}

\begin{document}

\thispagestyle{empty}

\begin{flushright}
AEI-2001-058\\
LPTENS-01/27
\end{flushright}
\renewcommand{\thefootnote}{\fnsymbol{footnote}}

\vspace*{0.1ex}
\begin{center}
\mathon
{\bf\LARGE $N\equ8$ matter coupled AdS$_3$ supergravities}
\mathoff
\bigskip\bigskip\medskip

{\bf H.~Nicolai\medskip\\ }
{\em Max-Planck-Institut f{\"u}r Gravitationsphysik,\\
  Albert-Einstein-Institut,
\footnote{Supported in part by the European Union under Contracts
No. HPRN-CT-2000-00122 and No. HPRN-CT-2000-00131.}\\
  M\"uhlenberg 1, D-14476 Golm, Germany}

\smallskip {\small nicolai@aei-potsdam.mpg.de} \bigskip\smallskip
\addtocounter{footnote}{-1}

{\bf H.~Samtleben\medskip\\ }
{\em Laboratoire de Physique Th{\'e}orique\\
  de l'{\'E}cole Normale Sup{\'e}rieure,
\footnotemark $^,$\footnote{UMR 8549: Unit{\'e} Mixte du Centre
National de la Recherche Scientifique, et de l'{\'E}cole Normale
Sup{\'e}rieure. }\\ 
24 Rue Lhomond, F-75231 Paris Cedex 05, France}~
\\

\smallskip {\small henning@lpt.ens.fr\medskip} 
\end{center}
\renewcommand{\thefootnote}{\arabic{footnote}}
\setcounter{footnote}{0}
\bigskip
\bigskip

\begin{abstract}
Following the recent construction of maximal ($N\equ16$) gauged 
supergravity in three dimensions, we derive gauged $D\equ3$, $N\equ8$ 
supergravities in three dimensions as deformations of the corresponding
ungauged theories with scalar manifolds $SO(8,n)/(SO(8)\cro SO(n))$. 
As a special case, we recover the $N\equ(4,4)$ theories with local 
$SO(4)=SO(3)_L\cro SO(3)_R$, which reproduce the symmetries and 
massless spectrum of $D\equ 6,N\equ(2,0)$ supergravity compactified
on AdS$_3 \cro S^3$.
\end{abstract}

\renewcommand{\thefootnote}{\arabic{footnote}}
\vfill

\bigskip

\leftline{{June 2001}}

\setcounter{footnote}{0}
\newpage

Gauged supergravities in three dimensions differ from the gauged
theories in higher dimensions by the existence of an on-shell duality
between the gauge fields and the scalar fields parametrizing the coset
manifolds (a subset of) whose isometries are gauged
\cite{NicSam01,NicSam01a}. For instance, to gauge the maximal
$N\equ16$ theory one starts from the maximally dualized theory in
which all bosonic physical degrees of freedom are contained in the
scalar coset space $G/H\equ E_{8(8)}/SO(16)$. Choosing a gauge group
$G_0\subset G$, the corresponding vector fields transforming in the
adjoint representation of $G_0$ are defined by duality as (nonlocal
and nonlinear) functions of the scalar fields, all of which are
kept. At the Lagrangian level the duality relation is implemented by
means of a Chern-Simons term in order of the gauge coupling constant
(rather than the usual Yang Mills term), such that the duality
equation relating the vectors to the scalar fields appears as an
equation of motion. As required by consistency, the vector fields then
do not introduce new physical degrees of freedom. A similar structure
has been found for the abelian gaugings of the $D\equ 3, N\equ2$
supergravities \cite{DKSS00}. While there are thus no a priori
restrictions on the choice of gauge group, the requirement that local
supersymmetry be preserved in the gauged theory poses strong
constraints. It was a main result of \cite{NicSam01,NicSam01a} that
all consistency conditions can be encoded into a single $G$-covariant
projection condition for the embedding tensor of $G_0$.

In this article we perform an analogous construction for the
nonmaximal $N\equ8$ theories of \cite{MarSch83} and show that the
techniques developed in \cite{NicSam01,NicSam01a} can be carried over
to this case, allowing for a quick determination of the possible gauge
groups. Pure (topological) $N\equ 8$ supergravity can be coupled to an
arbitrary number~$n$ of matter multiplets, each consisting of eight
bosonic and fermionic physical degrees of freedom. The scalar sectors
of these theories are governed by the coset manifolds~\cite{MarSch83}
\ben
G/H = SO(8,n)/(SO(8)\cro SO(n)) \;.
\een
Gauging any of these theories amounts to promoting a subgroup $G_0
\subset SO(8,n)$ to a local symmetry in such a way that the local
supersymmetry remains preserved. As for $N\equ 16$, the proof of
consistency can be reduced to a single projection condition for the
gauge group (see eq.~\Ref{crit} below), which is for instance solved
by the subgroups
\be
G_0~=~ SO(p,4\mis p) \times SO(q,4\mis q)  \;,
\qquad p, q = 0, 2 \;,
\la{ncI}
\ee
with independent gauge coupling constants $g_1$, $g_2$ for the two
factors in \Ref{ncI}. Hence, for each of the groups in \Ref{ncI}
there exists a gauged $N\equ8$ supergravity with local $G_0$ symmetry. 
There are further (non-compact) gauge groups which will be briefly
discussed at the end.

Half maximal gauged supergravities in three dimensions are expected to
describe e.g.~the massless sector of the AdS$_3 \cro S^3$ reduction of
$D\equ6$, $N\equ(2,0)$ supergravity coupled to $n_\scT$ tensor
multiplets \cite{Roma86,Ricc98}. Of particular interest
\cite{MalStr98,GiKuSe98} have been the theories with $n_\scT\equ 5$
and $n_\scT\equ 21$, which correspond to compactifying IIB
supergravity on AdS$_3 \cro S^3\cro M^4$ with $M^4 =T^4$ or $K3$,
respectively. The complete spectrum on AdS$_3 \cro S^3$ has been
computed in \cite{DKSS98,dBoe99}; it is organized by the supergroup
\be
SU(2|1,1)_L \times SU(2|1,1)_R \;,
\la{sug}
\ee
the $N\equ(4,4)$ extension of the AdS$_3$ group $SU(1,1)_L\cro
SU(1,1)_R$ containing the isometry group of the three sphere $SO(4)=
SO(3)_L \cro SO(3)_R$ which is the gauge symmetry of the theory. In
addition, the theory is invariant under a global $SO(4)$ which is a
remnant of the $R$-symmetry group $SO(5)\cong Sp(4)$ in six
dimensions. The massless sector has been found to consist of $n_\scT$
copies of the short $(2,2)_S$ multiplet of \Ref{sug} with eight
bosonic (scalar) and eight fermionic degrees of freedom (following
\cite{dBoe99}, the nomenclature $(k_1,k_2)_S$ for a multiplet of
\Ref{sug} refers to the representation content of its highest weight
state under $SO(4) \cong SO(3) \cro SO(3)$).  By contrast, the $SO(4)$
vector fields belong to the nonpropagating three-dimensional
supergravity multiplet. The low energy theory is then expected to
coincide with the gauged supergravity obtained by taking $p\equ
q\equ0$ in \Ref{ncI} and by setting $g_2=0$, such that the second
factor turns into a global $SO(4)$ symmetry while the associated
vector fields decouple from the Lagrangian. As we shall verify in more
detail below, this theory indeed reproduces the symmetries and
massless spectrum of the six-dimensional theory on AdS$_3 \cro S^3$.

\paragraph{The Lagrangian:}
The gauged theory is constructed by deforming the $N\equ8$ theory
of \cite{MarSch83} (to which we refer for notations) according
to the Noether procedure and by adding a Chern-Simons term for the 
vector fields transforming in the adjoint representation of the gauge
group. The scalar matter forms a $G/H=SO(8,n)/(SO(8)\cro SO(n))$ coset
space sigma model. It is most conveniently parametrized by $SO(8,n)$
valued matrices $L$ in the fundamental representation. Accordingly, 
the gauged Lagrangian is invariant under the symmetry
\be
L(x)\; \longrightarrow \; g_0 (x)\, L(x) \,h^{-1}(x)\;, \qquad
  g_0 (x) \in G_0 \; , \; h(x) \in H \;,
\label{GHV1}
\ee
where $G_0$ is an as yet undetermined subgroup of $G$. The choice of
$G_0$ turns out to be severely restricted by supersymmetry.  Upon
gauging, the global $G$ symmetry of the ungauged theory is broken down
to the centralizer of $G_0$ in $G$, which acts by left multiplication
on $L$. The scalar fields couple to the fermions via the
$G_0$-covariantized currents
\be
L^{-1}\left(\dd_\mu+g\GTh_{\cM\cN}\,B_\mu{}^\cM t^\cN\right) L ~=:~ 
\ft12\, \CQ_\mu^{IJ} \,X^{IJ} 
+ \ft12\, \CQ_\mu^{rs} X^{rs} +  \CP_\mu^{Ir} \,Y^{Ir}
\;,
\la{curG}
\ee
where $\{X^{IJ}, X^{rs}\}$ and $\{Y^{Ir}\}$ denote the compact and
noncompact generators of $\mathfrak{so}_{8,n}$, respectively, with
indices $I, J \in \underline{8}$, and $r, s \in \underline{n}$.  In
the following, we will use calligraphic letters as collective labels
for the entire set of generators of $\mathfrak{so}_{8,n}$: $\{t^\cM\}
= \{X^{IJ}, X^{rs},Y^{Ir}\}$; more specifically, we distinguish
between labels ${\CA, \CB,}\dots$ and ${\CM, \CN}, \dots$ to indicate
their transformation properties under the action of $H$ and
$G$, respectively, cf.~\Ref{V}.

The constant tensor $\Theta_{\cM\cN}$ characterizes the embedding of
the gauge group $G_0$ in $G$. It is obtained by restricting the
$\mathfrak{so}_{8,n}$ Cartan-Killing form $\eta_{\cM\cN}$ to the
simple factors of the algebra $\mfg_0\subset\mathfrak{so}_{8,n}$
associated to $G_0$, and is generally of the form
\be
\Theta_{\cM\cN} = \sum_j \varepsilon_j \, \eta^{(j)}_{\cM\cN} \;,
\la{emb}
\ee
where $\eta^{\cM\cN} \eta^{(j)}_{\cN\cK}$ project onto the simple
subfactors of $\mfg_0$, and the factors $\varepsilon_j$ correspond to
the relative coupling strengths. The Lagrangian of the gauged theory 
explicitly depends on the representation matrices $\VV{\cM}{\cA}$ 
in the adjoint representation of $SO(8,n)$
\be
L^{-1} t^\cM L ~\equiv~  \VV{\cM}{\cA} \,t^\cA ~=~ 
\ft12\,\VV{\cM}{IJ}\,X^{IJ} 
+ \ft12\, \VV{\cM}{rs} \,X^{rs} +  \VV{\cM}{Ir} \,Y^{Ir}
\;,
\la{V}
\ee
via the $T$-tensor
\be
\TT{\cA}{\cB} ~\equiv~ \GTh_{\cM\cN}\,\VV{\cM}{\cA}\VV{\cN}{\cB} \;.
\la{T}
\ee

The construction of the gauged theory parallels the one of the maximal
theory \cite{NicSam01,NicSam01a} and we simply state the resulting
Lagrangian (up to quartic fermionic terms)
\ba
\CL&=& -\ft14 \,eR
+\ft12 \Ge^{\mu\nu\rho} \Bpsi^A_\mu D_\nu\psi^A_\rho
+ \ft14\,e \CP_\mu^{Ir} \CP^\mu{}^{\,Ir} 
-\ft12\,\I e \Bchi^{\dA r}\Gg^\mu D_\mu\chi^{\dA r} 
\non[1ex]
&&{}
-\ft14\,\Ge^{\mu\nu\rho}\,g\GTh_{\cM\cN}\,B_\mu{}^\cM
\Big(\dd_\nu B_\rho\,{}^\cN
+\ft13\,g \GTh_{\cK\cL}\, f^{\cN\cK}{}_{\cP}\,
B_\nu{}^\cL B_\rho{}^\cP \Big) 
\non
&&{}
- \ft12\,e \CP_\mu^{Ir}\Bchi^{\dA r} \GG^I_{A\dA}\Gg^\nu\Gg^\mu\psi^A_\nu 
+\ft12\,ge A_1^{AB} \,\Bpsi{}^A_\mu \Gg^{\mu\nu} \psi^B_\nu
\non[1ex]
&&{}
+\I\,ge A_2^{A\dA r} \,\Bchi^{\dA r} \Gg^{\mu} \psi^A_\mu
+\ft12\,ge A_3^{\dA r\, \dB s}\, \Bchi^{\dA r}\chi^{\dB s} 
+ e\, W
\;.
\la{L}
\ea
Gravitini and matter fermions transform in the spinor and conjugate
spinor representations of $SO(8)$, denoted by indices $\indA,\indB,
\dots$, and $\inddA, \inddB, \dots$,
respectively.\footnote{Generically, the supersymmetry parameters and
gravitini belong to the vector representation of $SO(N)$ for
$N$-extended supergravity. In order to facilitate the comparison with
\cite{MarSch83} we here adopt an assignment of representations which
differs from the generic one by a triality rotation.} Their
covariant derivatives are built with the $H$-currents $\CQ_\mu$ from
\Ref{curG}:
\ba
D^{\vphantom b}_\mu \psi_\nu^A &:=& 
\dd^{\vphantom b}_\mu \psi_\nu^A + 
\ft14\,\Go_\mu{}^{ab}\,\Gg_{ab}\, \psi_\nu^A  +
\ft14\,\CQ_\mu^{IJ}\GG^{IJ}_{AB} \psi_\nu^B 
\;,\non
D_\mu \chi^{\dA r} &:=& \dd_\mu \chi^{\dA r} + 
\ft14\,\Go_\mu{}^{ab}\,\Gg_{ab}\, \chi^{\dA r}  +
\ft14\, \CQ_\mu^{IJ} \GG^{IJ}_{\dA\dB} \,\chi^{\dB r} +
\CQ_\mu^{rs} \,\chi^{\dA s} \;.
\nn
\ea
The scalar tensors $A_{1,2,3}$ describing the Yukawa-like coupling
between scalars and fermions may be expressed as linear functions of
the $T$-tensor \Ref{T}
\ba A_1^{AB}&=& -\Gd^{AB}\,\Gth -
\ft1{48}\,\GG^{IJKL}_{AB}\,\TT{IJ}{KL} \;,\non A_2^{A\dA r}&=&
-\ft1{12}\,\GG^{IJK}_{A\dA}\,\TT{IJ}{Kr} \;,\non A_3^{\dA r\, \dB
s}&=& 2\Gd^{\dA\dB}\Gd^{rs}\,\Gth +
\ft1{48}\,\Gd^{rs}\,\GG^{IJKL}_{\dA\dB}\,\TT{IJ}{KL}
+\ft12\,\GG^{IJ}_{\dA\dB}\,\TT{IJ}{rs} \;, \la{A123} \ea
where by $\Gth\!\equiv\!\ft2{(8\pls n)(7\pls n)}\,
\eta^{\cM\cN}\GTh_{\cM\cN}$ we denote the trace of \Ref{emb} which
equals the trace of the $T$-tensor. The scalar potential $W$ is
given by the following expression (of eighth order in the matrix
entries of $L$):
\be
W~=~\ft14\,g^2
\left(A_1^{AB}A_1^{AB}-\ft12\,A_2^{A\dA r}A_2^{A\dA r}\right) \;.
\ee
The supersymmetry variations are
\ben
\begin{array}{rclrcl}
L^{-1}\Gd L &\equ& Y^{Ir}\;\Beps^A\,\GG^I_{A\dA}\chi^{\dA r}\;, &\quad\;
\Gd\chi^{\dA r} &\equ& \ft12\I\GG^I_{A\dA}\Gg^\mu\Ge^A \CP_\mu^{Ir}
+g A_2^{A\dA r}\Ge^A \;, \\[1ex]
\Gd e_\mu{}^\Ga &\equ& \I \Beps^A\,\Gg^\Ga\psi^A_\mu \;, &
\Gd\psi^A_\mu &\equ& D_\mu \Ge^A + \I g A_1^{AB}\Gg_\mu\Ge^B \;, \\[1ex]
\Gd B_\mu{}^\cM &\equ& 
\multicolumn{4}{l}{-\ft12\VV{\cM}{IJ}\,\Beps^A\,\GG^{IJ}_{AB}\psi^B_\mu
+ \I\,\VV{\cM}{Ir}\,\Beps^A\,\GG^I_{A\dA}\Gg_\mu\chi^{\dA r} \;.}
\end{array}
\een
Lagrangian and supersymmetry transformations have been given up to
higher order fermionic terms which were already given in \cite{MarSch83}
(as shown in \cite{NicSam01a} the gauging does not lead to any 
modification in these terms). As in the maximal gauged theory, 
invariance of the Lagrangian \Ref{L} under these transformations 
implies several consistency conditions on the tensors $A_{1,2,3}$, 
which are solved by \Ref{A123} provided that the $T$-tensor 
satisfies certain identities. Combining these identities into 
$SO(8,n)$ covariant expressions, we obtain an identity for the 
embedding tensor $\GTh_{\cM\cN}$ which allows us to select
the admissible gauge groups $G_0$.

\paragraph{\mathon Group theory and $T$-identities: \mathoff}

The embedding tensor $\GTh_{\cM\cN}$ is a $G_0$-invariant tensor in
the symmetric tensor product of two adjoint representations of
$SO(8,n)$. It may accordingly be decomposed into its irreducible parts
\be
\GTh_{\cM\cN} ~\;\subset\;~ \left( \; 
\young(\hfil,\hfil) \;\times \, \young(\hfil,\hfil)\; \right)_{\rm sym} 
~\;=\;~ 1 \;\;+\;\;
\young(\hfil\hfil) \;\;+\;\;
\young(\hfil\hfil,\hfil\hfil) 
\;\;+\;\; \yng(1,1,1,1) \; ,
\la{TDEC}
\ee
where each box represents a vector representation \vecso~of
$SO(8,n)$. We here use the standard Young tableaux for the orthogonal
groups: for instance,\ {\tiny \young(\hfil\hfil)} denotes the
traceless part of the symmetric tensor product
$\big($\vecso$\times$\vecso $\big)_{\rm sym}$. The four
irreducible parts in the decomposition \Ref{TDEC} have dimensions 1,
$\ft12 (8\pls n)(9\pls n)-1$, $\ft1{12}(5\pls n)(8\pls n)(9\pls
n)(10\pls n)$, and ${8\pls n}\choose{4}$, respectively.
Under $H=SO(8)\cro SO(n)$ the vector representation decomposes as
\ben
\young(\hfil) ~\ra~ (8_v,1) + (1,\mbox{\tiny \young(\bullet)}\,)
\;\;,
\een
where by {\tiny \young(\bullet)} we now denote the vector
representation $n$ of $SO(n)$. The $T$-tensor which according to
\Ref{T} is obtained by an $SO(8,n)$ rotation with $\cV$ from
\Ref{TDEC}, accordingly decomposes as
\ba
\TT{IJ}{KL} &=& (1,1) + (35_s,1) + (35_c,1) + (35_v,1) + (300,1) 
\;,
\non
\TT{IJ}{rs} &=& (28,\mbox{\tiny $\young(\bullet,\bullet)$}\;) 
\;,
\non
\TT{rs}{pq} &=& (1,1)
+ (1,\mbox{\tiny \young(\bullet\bullet)})
+ (1,\mbox{\tiny $\young(\bullet\bullet,\bullet\bullet)$}\;) 
+ (1,\mbox{\tiny $\young(\bullet,\bullet,\bullet,\bullet)$}\; )
\;,
\non
\TT{IJ}{Kp} &=& (8_v,\mbox{\tiny \young(\bullet)}\,)
+(56_v,\mbox{\tiny \young(\bullet)}\,)
+(160_v,\mbox{\tiny \young(\bullet)}\,)
\;,
\non
\TT{Kp}{rs} &=& (8_v,\mbox{\tiny \young(\bullet)}\; )
+ (8_v,\mbox{\tiny $\young(\bullet\bullet,\bullet)$}\;)
+ (8_v,\mbox{\tiny $\young(\bullet,\bullet,\bullet)$}\; )
\;,
\non
\TT{Ir}{Js} &=& (1,1) + (1,\mbox{\tiny \young(\bullet\bullet)}\;) + 
(35_v,1) + (35_v,\mbox{\tiny \young(\bullet\bullet)}\;) 
+ (28,\mbox{\tiny $\young(\bullet,\bullet)$}\;)
\;,
\la{decT}
\ea
under $SO(8)\cro SO(n)$. As in the maximal gauged theory
\cite{NicSam01,NicSam01a} it turns out that the supersymmetry of the
gauged Lagrangian \Ref{L} is equivalent to the vanishing of certain
subrepresentations in \Ref{decT}, whereas the nonvanishing parts may
be cast into a single $SO(8,n)$ representation. The relevant
components of the $T$-tensor may be entirely expressed in terms 
of the scalar tensors $A_{1,2,3}$
\ba
\TT{IJ}{KL} &=& -\Gth\,\Gd^{IJ}_{KL} 
-\ft18\,\GG^{IJKL}_{AB}\,A_1^{AB} 
+\ft1{8n}\,\GG^{IJKL}_{\dA \dB}\,A_3^{\dA r\,\dB s} 
\;,
\non
\TT{IJ}{rs} &=& \ft18\,\GG^{IJ}_{\dA\dB}\,A_3^{\dA r\,\dB s}
\;,
\non
\TT{IJ}{Kp} &=& -\ft14\,\GG^{IJK}_{A\dA}\,A_2^{A\dA r}
\;,
\non
\TT{Ir}{Js} &=& \Gth\,\Gd_{IJ}\Gd_{rs} + 
\ft18\,\GG^{IJ}_{\dA\dB}\,A_3^{\dA r\,\dB s}
\;.
\la{TinA}
\ea
Comparing this to the $SO(8)\cro SO(n)$ representation content of the
Yukawa tensors $A_{1,2,3}$ extracted from \Ref{A123}
\ba
A_1^{AB} &=& (1,1) + (35_s,1) 
\;,
\non
A_2^{A\dA r} &=& (56_v,\mbox{\tiny \young(\bullet)}) 
\;,
\non
A_3^{\dA r\,\dB s} &=& 
(1,1) + 
(35_c,1) 
+ (28,\mbox{\tiny $\young(\bullet,\bullet)$}\;) 
\;,
\nn
\ea
we recognize that, apart from the singlet contributions, the nonvanishing 
part of $T_{\cA|\cB}$ precisely corresponds to the last term in
\Ref{TDEC} which decomposes into
\ben
\yng(1,1,1,1) ~=~ (35_s,1) + (35_c,1) 
+ (56_v,\mbox{\tiny \young(\bullet)}\,)
+ (28,\mbox{\tiny $\young(\bullet,\bullet)$}\;) 
+ (8_v,\mbox{\tiny $\young(\bullet,\bullet,\bullet)$}\; )
+ (1,\mbox{\tiny $\young(\bullet,\bullet,\bullet,\bullet)$}\; ) \;.
\een
Together, we obtain the $SO(8,n)$ covariant consistency criterion
\Yboxdim3pt
\ba
\TT{\cA}{\cB} &=& \Gth \,\eta_{\cA\cB} +
{\mathbb P}\!_{\atop{}{\yng(1,1,1,1)}}\,_{\cA\cB}{}^{\cC\cD}\;
\TT{\cC}{\cD}
\non[1ex]
&&{}
\Longleftrightarrow \qquad
\GTh_{\cA\cB} ~=~ \Gth \,\eta_{\cA\cB} +
{\mathbb P}\!_{\atop{}{\yng(1,1,1,1)}}\,_{\cA\cB}{}^{\cC\cD}\;
\GTh_{\cC\cD}
\;\;,
\la{crit}
\ea
for $T_{\cA|\cB}$ or equivalently for the embedding tensor
$\GTh_{\cM\cN}$ of the gauge group, where ${\mathbb
P}\!_{\atop{}{\yng(1,1,1,1)}}$ denotes the projector onto the
corresponding part in the tensor product \Ref{TDEC}. In components,
this condition reads
\be
\begin{array}{rclrcl}
\GTh_{IJ,KL} &=& -\Gth\,\Gd^{IJ}_{KL} + \GTh_{[IJ,KL]} \;,
&\qquad
\GTh_{IJ,Kp} &=& \GTh_{[IJ,K]p}
\;,
\\[1ex]
\GTh_{rs,pq} &=& -\Gth\,\Gd^{rs}_{pq} + \GTh_{[rs,pq]} \;,
&
\GTh_{Kp,rs} &=& \GTh_{K[p,rs]}
\;,
\\[1ex]
\GTh_{Ir,Js} &=& 
\multicolumn{4}{l}{\Gth\,\Gd_{IJ}\Gd_{rs} + \GTh_{IJ,rs}\;,}
\end{array}
\la{critTh}
\ee
and likewise for $T_{\cA|\cB}$, in agreement with \Ref{TinA}. {}From
the above decomposition one sees that $T_{\cA|\cB}$ possesses the
additional nonvanishing components
\ben
\TT{rs}{pq} ~=~ -\Gth\,\Gd^{rs}_{pq} + \TT{[rs}{pq]} \;,
\qquad
\TT{Kp}{rs} ~=~ \TT{K[p}{rs]} \;,
\een
which do not appear in the Lagrangian. In turn, it may be verified 
by lengthy but straightforward computation that \Ref{critTh}
(together with the fact that $\GTh_{\cM\cN}$ projects onto a subgroup)
indeed encodes the full set of identities which are required for
supersymmetry of the gauged Lagrangian \Ref{L}.

\paragraph{Admissible gauge groups:} 
It remains to solve \Ref{crit} and to identify the subgroups of
$SO(8,n)$ whose embedding tensor satisfies the projection constraint
\Ref{crit} such that the $G_0$-gauged Lagrangian \Ref{L} remains
supersymmetric. Rather than aiming for a complete classification, we
here wish to discuss the most interesting cases, and, in particular,
the gauging of the compact $SO(4)\subset SO(8)$ related to the
six-dimensional supergravity on AdS$_3\cro S^3$.

Compact gauge groups $G_0\subset SO(8)\cro SO(n)$ satisfy
$\GTh_{Ir,Js}=0$; according to \Ref{critTh} they hence factor into two
subgroups of $SO(8)$ and $SO(n)$, respectively, such that each factor
separately satisfies \Ref{crit}. It is straightforward to show that
none of the maximal subgroups $SO(p)\cro SO(8\!-\!p)\subset SO(8)$ for
$p\not=4$ satisfies \Ref{crit}. The case $p\equ4$ requires a separate
analysis since the two factors $SO(4)\cro SO(4)$ are not simple but
both factor into $SO(4)=SO(3)_L\cro SO(3)_R$. We therefore consider
the group
\ba
G_0 &=& SO(4)^{(1)}\cro SO(4)^{(2)} \non
&=&
\left( SO(3)^{(1)}_L\cro SO(3)^{(1)}_R\right) \times 
\left( SO(3)^{(2)}_L\cro SO(3)^{(2)}_R\right) 
\;.
\la{G0}
\ea 
Denote by $I=\{i,\ovi\,\}$ the corresponding decomposition of the
$SO(8)$ vector indices $8\ra (4,1) + (1,4)$. The embedding tensors of
the four simple factors of $G_0$ are given by
\be
\eta^{(1\pm)}_{IJ,KL} ~=~ 
\left(\Gd_{ij}^{kl}\pm\ft12\Ge^\vl_{ijkl}\right)
\;,\qquad
\eta^{(2\pm)}_{IJ,KL} ~=~ 
\left(\Gd_{\ovi\ovj}^{\ovk\ovl}\pm\ft12\Ge^\vl_{\ovi\ovj\ovk\ovl} \right) 
\;.
\la{eta}
\ee
The Cartan-Killing form of $SO(8)$ e.g.\ decomposes as
\ben
\eta^\vl_{IJ,KL} ~=~ \eta^{(1+)}_{IJ,KL} + \eta^{(1-)}_{IJ,KL} +
\eta^{(2+)}_{IJ,KL} + \eta^{(2-)}_{IJ,KL} + \dots  \;.
\een
The condition \Ref{crit} is obviously solved by the following linear
combination of the simple embedding tensors
\ba
\GTh_{IJ,KL} &=&
\left(\eta^{(1+)}_{IJ,KL} - \eta^{(1-)}_{IJ,KL}\right) + 
\alpha \left(\eta^{(2+)}_{IJ,KL} - \eta^{(2-)}_{IJ,KL}\right)
\non[1ex]
&=& \Ge^\vl_{ijkl} + \alpha\,\Ge^\vl_{\ovi\ovj\ovk\ovl}  \;,
\la{so4so4}
\ea
with a free constant $\Ga$. The trace part of $\GTh$ vanishes. This is
the embedding tensor of \Ref{G0} where the relative coupling constants
of the two factors in each $SO(4)$ differ by a factor of $-1$ whereas
the relative coupling constant $\Ga$ between the two $SO(4)$ factors
may be chosen arbitrarily. The resulting theory \Ref{L} has gauge
group $SO(4)\cro SO(4)$ with two independent gauge coupling constants.

As a special case, we can choose $\Ga\equ0$ in \Ref{so4so4} and 
obtain a theory with gauge group
\be 
G_0~=~SO(4)~=~SO(3)_L\cro SO(3)_R \;.  
\la{G00}
\ee
In this case, the second $SO(4)$ factor of \Ref{G0} survives as a
global symmetry of the theory.  As explained above, the symmetries of
this theory thus coincide with those of $D\equ6$, $N\equ(2,0)$
supergravity on AdS$_3 \cro S^3$.  The fields of \Ref{L} accordingly
decompose under \Ref{G00} into the physical degrees of freedom
contained in $n$ matter supermultiplets\footnote{The multiplicities
$1$, $2_\pm$, $4$, here and in the following formula refer to
irreducible representations of the second (global) SO(4) symmetry.}
\be
n \Big( 1\!\cdot\!(2,2)+4\!\cdot\!(1,1)
+2_+\!\cdot\!(2,1)+2_-\!\cdot\!(1,2) \Big)
\;, 
\la{fc}
\ee
and the nonpropagating fields which include the graviton, gravitini
and the $SO(4)$ Chern-Simons vector fields.  Dropping the physical
fields \Ref{fc} from the Lagrangian, one recovers one of the
topological Chern-Simons supergravities of \cite{AchTow86}.

The theories obtained with \Ref{so4so4} have a maximally
supersymmetric ground state at $L\equ I$, i.e.\ for vanishing scalar
fields. Its background isometries form the product of simple
supergroups
\be
D^1(2,1;\alpha)_L \times D^1(2,1;\alpha)_R \;,
\la{isom}
\ee
which is an $N\equ(4,4)$ extension of the AdS$_3$ group $SU(1,1)_L\cro
SU(1,1)_R$~\cite{GuSiTo86}. We would expect the corresponding gauged
supergravity \Ref{L} to be related to the ${\rm AdS}_3\cro S^3\cro
S^3$ compactifications considered in \cite{BoPeSk98,dBPaSk99}.  For
$\alpha\equ 0$, i.e.\ for the theory with gauge group \Ref{G00}, the
background isometry group \Ref{isom} factors into the semi-direct
product of the $N\equ(4,4)$ supergroup
\be
SU(2|1,1)_L \times SU(2|1,1)_R \;,
\la{isom0}
\ee
and the global $SO(4)$ symmetry of the gauged theory. The ground state
$L\equ I$ of this theory corresponds to the AdS$_3\cro S^3$ vacuum of
the six-dimensional theory.  The physical field content \Ref{fc}
corresponds to $n$ copies of the $(2,2)_S$ short multiplet of
\Ref{isom0}, whereas the nonpropagating fields combine into the
$(3,1)_S\pls(1,3)_S$ short multiplets of \Ref{isom0}. As anticipated
above, this reproduces the massless spectrum of $D\equ 6$, $N\equ
(2,0)$ supergravity with $n$ tensor multiplets on AdS$_3 \cro S^3$
\cite{DKSS98,dBoe99}.

Similarly, we may gauge subgroups of the compact $SO(n)$ by embedding
up to $[n/4]$ factors of $SO(4)=SO(3)\cro SO(3)$, each of which is
twisted by a relative factor of $-1$ between the simple embedding
tensors of its two $SO(3)$ factors. This allows to introduce up to
$[n/4]$ additional independent coupling constants.

The noncompact gaugings in \Ref{ncI} are obtained in the same way. By
replacing some of the $SO(8)$ vector indices $I, J, \dots$ in
\Ref{so4so4} by $SO(n)$ vector indices $r, s, \dots$, one may embed
\be
G_0~=~ SO(p,4\mis p) \cro SO(q,4\mis q)  \;,
\la{nc}
\ee
where $p$ and $q$ may take the values $0$ or $2$, and arbitrary
relative coupling constant $\Ga$ between the two factors in
\Ref{nc}. The consistency of these theories follows in complete
analogy to the compact case from the projection form of the criterion
\Ref{crit}. By contrast, the noncompact group $SO(3,1)\cong
SL(2,{\mathbb C})$ cannot be consistently gauged because the
$\epsilon$-tensor in \Ref{eta} would have to carry a factor of $i$,
resulting in an imaginary gauge coupling constant.

In addition to these gauge groups, the consistency condition
\Ref{crit} allows for several other noncompact maximal subgroups of
$G\equ SO(8,n)$.  These solutions may be found by group theoretical
arguments analogous to the ones used in \cite{NicSam01,NicSam01a} to
derive the exceptional gauging. E.g.\ for $n\equ8$, the group
$G\!=\!SO(8,8)$ possesses the maximal noncompact subgroup
\be
G_0 ~=~ G^{(1)}\cro G^{(2)} ~=~ C_{4(4)}\!\times\!SL(2,\mathbb{R})
\;.
\la{C44}
\ee
whose compact subgroup $H_0=H^{(1)}\cro H^{(2)}=\left(SU(4) \cro
SO(2)\right)\times SO(2)$ is embedded in $H=SO(8)\cro SO(8)$ via the
maximal embeddings $SO(8)\supset SO(6)\cro SO(2)$, where the
$SU(4)\!\sim\!SO(6)$ factor of $H_0$ lies in the diagonal. The
embedding tensor of \Ref{C44} with a fixed particular ratio of the
coupling constants satisfies \Ref{crit} where this ratio is obtained
from the formula \cite{NicSam01a}
\be
\frac{g_1}{g_2} ~=~ 
-\frac{7\, {\rm dim\,} G^{(2)}-15\, {\rm dim\,} H^{(2)}}
{7\, {\rm dim\,} G^{(1)}-15\, {\rm dim\,} H^{(1)}} ~=~ -\frac12
\;\;.
\la{ratio}
\ee
The same argument holds for the other real form of this gauge group
\be
G_0 = C_{4(-4)}\!\times\!SO(3) \;,
\la{C4m4}
\ee
contained in $G$. Its maximal compact subgroup $H_0=\left(U\!Sp(4)\cro
U\!Sp(4)\right)\cro SO(3)$ is embedded in $H$ via the maximal
embeddings $SO(8)\supset SO(5)\cro SO(3)$, where the $SO(3)$ factor of
$H_0$ lies in the diagonal. The ratio of coupling constants is again
obtained from \Ref{ratio} and gives the same value for $g_1/g_2= -1/2$.
The latter construction \Ref{C4m4} may be further generalized to any
$n\in 4\,{\mathbb Z}$: for $G=SO(8,4m)$, the maximal noncompact
subgroup
\be
G_0 = Sp(m,2)\cro SO(3) \;,
\ee
satisfies \Ref{crit} if the ratio of coupling constants is given by
$g_1/g_2= -2/(2\pls m)$. It would be interesting to identify a possible
higher dimensional origin of these noncompact gaugings. 

Finally, one can expect that other gauged theories with $N\!<\!16$
local supersymmetries can be constructed in a similar fashion (a
complete list of extended supergravities in three dimensions has been
given in \cite{dWToNi93}).


\providecommand{\href}[2]{#2}\begingroup\raggedright\endgroup

\end{document}